\documentclass{sig-alternate}

\usepackage{balance}
\usepackage{wasysym}
\usepackage{booktabs} 
\usepackage{cite}
\usepackage{color}
\usepackage{stackrel}
\usepackage{siunitx}
\usepackage{xspace}   
\usepackage{enumitem} 
\usepackage{url}
\usepackage{graphicx}
\usepackage{adjustbox}
\newcolumntype{R}[2]{%
    >{\adjustbox{angle=#1,lap=\width-(#2)}\bgroup}%
    l%
    <{\egroup}%
}

\pdfoutput=1
\newcommand{\Paragraph}[1]{\smallskip\noindent{\bf #1.}}

\newcommand*\rota{\multicolumn{1}{R{90}{0.5em}} }

\newcommand{\Tdot}{$\CIRCLE$}

\makeatletter
\def\@copyrightspace{\relax}
\makeatother

\def\sharedaffiliation{%
\end{tabular}
\begin{tabular}{c}}

\begin{document}

\title{IoTScanner: Detecting and Classifying
  Privacy Threats in IoT Neighborhoods}

\numberofauthors{3}
 \author{
 \alignauthor Sandra Siby\\
 \email{sandra\_ds\\@sutd.edu.sg}
 \alignauthor Rajib Ranjan Maiti\\
 \email{rajib\_maiti\\@sutd.edu.sg}\\
 \alignauthor Nils Ole Tippenhauer\\
 \email{nils\_tippenhauer\\@sutd.edu.sg}\\
 \sharedaffiliation
 \affaddr{Singapore University of Technology and Design (SUTD)}\\
 \affaddr{8 Somapah Road}\\
 \affaddr{487372 Singapore}\\
 }

\maketitle

\begin{abstract}

  In the context of the emerging Internet of Things (IoT), a
  proliferation of wireless connectivity can be expected. That
  ubiquitous wireless communication will be hard to centrally manage
  and control, and can be expected to be opaque to end users. As a
  result, owners and users of physical space are threatened to lose control 
  over their digital environments.
		
  In this work, we propose the idea of an IoTScanner. The IoTScanner
  integrates a range of radios to allow local reconnaissance of
  existing wireless infrastructure and participating nodes. 
  It enumerates such devices, identifies connection patterns, and provides
  valuable insights for technical support and home users alike. 
  Using our IoTScanner, we attempt to classify actively streaming IP cameras from 
  other non-camera devices using simple heuristics.
  We show that our classification approach achieves a high accuracy 
  in an IoT setting consisting of a large number of IoT devices. 
  While related work usually focuses on detecting either the infrastructure, or
  eavesdropping on traffic from a specific node, we focus on providing
  a general overview of operations in all observed networks. We do not
  assume prior knowledge of used SSIDs, preshared passwords, or
  similar.
	
\end{abstract}

\section{Introduction}
\label{sec:intro}

In the context of the emerging Internet of Things (IoT), a
proliferation of wireless connectivity can be expected, with
predictions reaching as many as 500 smart devices per household in
2022~\cite{gartner15iot}.  Heterogeneous smart devices that require
transparent Internet connectivity need to be integrated into a common
infrastructure.  In particular, communication standards such as
Zigbee, Bluetooth, Bluetooth Low Energy, and WiFi are expected to
provide such infrastructure either as mesh networks, or traditional
single-hop access points.  Often, vendors will sell their own
application gateway with integrated access point, which to a certain
degree hides the underlying communication from the owner.  Although
wireless communications have many benefits in terms of usability,
flexibility, and accessibility, there are also security concerns~\cite{sicari2015security}.
Among those concerns are privacy and in general, controllability. With
this growth in the size and complexity of wireless networks,
appropriate tools are required to better understand the environment's
wireless traffic.

In this work, we propose the idea of an IoTScanner. The IoTScanner is
a system that allows for passive, real-time monitoring of an existing
wireless infrastructure. The IoTScanner classifies and identifies
devices that are communicating using the infrastructure and traffic
patterns among the participating devices.  The IoTScanner will supply
this information in a structured manner so as to provide valuable
insights for technical support and home users alike.

While related work usually focuses on detecting either the
infrastructure, or eavesdropping on traffic from a specific node, we
focus on providing a general overview of operations in all observed
networks. We do not assume prior knowledge of used SSIDs, preshared
passwords, or similar.  In addition to this, the IoTScanner operates
passively, without active probing or other transmission.  Our emphasis
is also on providing real-time analysis and visualization of the
scanned network, unlike some related work that focuses on offline
analysis.

We note that we do not consider physical layer effects such as
collisions and packet loss, goodput, or interference between networks
in this work. We leverage recent developments in the area of cheap
software-defined radio modules to handle the physical layer of the
wireless traffic to provide the Link layer traffic. As most wireless
communication standards nowadays by default encrypt the Link layer
payload, we consider only Link layer traffic for analysis in this work
(without considering higher layer traffic). In particular, we aim to address the issue of providing controllability and privacy to the user of an IoT environment. The IoTScanner can provide details on the devices, the links between them and the amount of traffic in any environment in which it is placed. This information can then be used to classify and identify devices that could impact the privacy of the user. Three communication standards (WiFi, Bluetooth Low Energy (BLE) and Zigbee) are considered in this work.

We summarize our contributions as follows:
\begin{itemize}
\item We identify the issue of \emph{opaqueness} in IoTs:integration
  of heterogeneous IoT devices will lead to a plethora of wireless
  standard and networks operated in parallel.  Without a system like
  the IoT scanner, it will be difficult to keep overview of
  communication in a space (in particular for third-party networks).
\item We propose an abstract system design that allows to seamlessly
  integrate a range of radio devices with a scanning server.  The
  server's data can then be accessed by users through a RESTful
  API. The data obtained via the REST API can then be used for
  visualization of the scanned environment and further analysis.
\item We present an implementation of the proposed design, and demonstrate its effectiveness through a set of experiments.
\item We discuss how our system can be potentially used to identify privacy threats in an IoT environment.
\end{itemize}

The structure of this work is as follows: In
Section~\ref{sec:background}, we briefly summarize the relevant
background for our work. We propose our system design for the IoTScanner in Section~\ref{sec:scanner}. Our implementation is described
in Section~\ref{sec:implementation}. In Section~\ref{sec:wifi_exp},
we run some experiments to evaluate our implementation in an environment that consists of WiFi devices. In addition, we discuss the feasibility of the IoTScanner to perform traffic classification and device identification in this section. Sections ~\ref{sec:btevaluation} and ~\ref{sec:zigevaluation} describe experiments in BLE and Zigbee environments. We discuss some of the challenges we faced in Section ~\ref{sec:discussion}. Related work and comparison of features with existing tools
is summarized in Section~\ref{sec:related}.  We conclude the paper and
discuss possible future work in Section~\ref{sec:conclusions}.
  
\section{Background}
\label{sec:background}

\subsection{Internet of Things}
The Internet of Things (IoT) refers to a network of physical devices
that have the capability of gathering and transferring data from the
environment, and interacting with remote computers over Internet~\cite{atzori2010internet,gubbi2013internet}. 
IoT devices can be anything from cellphones to smart lamps, as long as
they have the ability to transfer data over the Internet, which
essentially means they are assigned some public IP addresses.

\subsection{Passive Monitoring}
Passive monitoring is a technique for observing the traffic in a
network only by listening to signals that are already available in the
network, without injecting any extra signal.  Unlike active
monitoring, passive monitors do not probe the devices they are
observing ~\cite{cottrell2001passive}.  In the context of wireless network, the technique is more
suitable as it only requires a traffic interceptor to be physically
present in the environment, without requiring any wire tapping or
so. Devices called sniffers or monitors are placed in the network to
intercept frames transmitted by devices in their vicinity.  Radios
that can be used to sniff different standards (WiFi, Bluetooth,
Zigbee) are available on the market and can be used for passive
monitoring.  

\subsection{System Model}
\label{sec:model}

In our systems model, we assume that there are one or more wireless
networks in an IoT scenario in which the scanner is placed.  For
example, in a smart home the electronic gadgets like smart lighting
can make use of Zigbee communication, personalized health monitoring
can use Bluetooth, Internet access can happen via WiFi communication.
The scenario can have one or more IoT devices, but not all the devices
need to be active (i.e., sending or receiving some information) all
the time.  For example, a smart blood pressure monitor may get
activated only about 2-3 times a day, whereas smart light need to be
active throughout the day.  In the basic operation mode, the scanner
needs no prior knowledge of the infrastructure and network
configuration in the IoT environment, i.e, no information is required
about which device used for what purpose, or what kind of encryption
is used by them, if any.  We assume that the user of our IoTScanner
system does not have any control over the devices, or know if they are
present or where exactly they are located.  Hence, there is a chance
that IoT devices that are present in the environment (but not actively
participating in the network) are not detected by the
scanner.  

\subsection{Attacker Model}
We assume a limited attacker who is \emph{honest but curious}. For
example, the attacker might access a webcam set up in a room to spy on
persons in the room, but the attacker will not set up a dedicated
device for this (in particular, the device will not use some
non-standard wireless communication). By obtaining the images from the
camera over a certain time period, the attacker violates the privacy
of the user. How the attacker obtained access to the camera does not
matter for our analysis.

As we do not assume that we have control over the network(s) in the
environment, we will not be able to detect the camera activity on the Network Layer or at a gateway or similar.

\section{A Passive Analysis Framework}
\label{sec:scanner}

In the following we provide details on our proposed passive analysis framework. We start with a concise problem statement, summarize the intuition behind our design and then provide additional details on individual components.
\subsection{Problem Statement}
\label{sec:statement}

In this work, we address the following problem: \emph{``In an IoT
  environment, which devices are present, and communicating via
  wireless communications?''} That information can then be used to
address questions such as: \emph{``Are the IoT devices in my environment used by an attacker to violate my privacy?''}  Ideally, such a goal should be
reached passively, without actively interfering with the environment
(which could be a public place such as an airport, hotel, or similar).

This problem cannot be solved by normal end-users easily; it requires specific software, hardware, and technical understanding of wireless protocols. We refer to this challenge as \emph{wireless opaqueness} for the end users: they are agnostic to the way networking works, which links exists, and how data flows in the neighborhood.

\subsection{Intuition}
To address the problem stated above, we now propose the
\emph{IoTScanner}. The aim of the IoTScanner is to provide real-time,
passive monitoring of an existing wireless infrastructure that
potentially constitute an IoT environment.  The scanner will identify
active devices that are communicating using that infrastructure, and
attempt to categorize IoT traffic depending on features such as the volume of
traffic observed. 

In particular, we use to following constraints to achieve passive monitoring. The IoTScanner:
\begin{itemize}
	\item will not associate with any access point present,
	\item will not perform active probing or fingerprinting, and
	\item will not decrypt the observed network traffic.
\end{itemize}
The IoTScanner only observes the network traffic at the Link layer, and then analyzes this traffic using frame header information. A more offensive active scanner would not need to follow those constraints.

The long term goal is to turn the IoTScanner into a convenient
hand-held device. In the context of this paper, we are using a
Raspberry Pi 3~\cite{raspberry} as platform, together with devices
such as Android tablets for visualization via a web application.

\subsection{IoTScanner: System Architecture}
The main features provided by our IoTScanner are
ubiquitous (wireless) signal interception, packet filtering, analysis
of the captured packets, and storage of the results. Finally, results
will be accessible via APIs to be visualized through a web-based 
visualization system. 

\begin{figure}[tb]
	\centering
	\includegraphics[width=\linewidth]{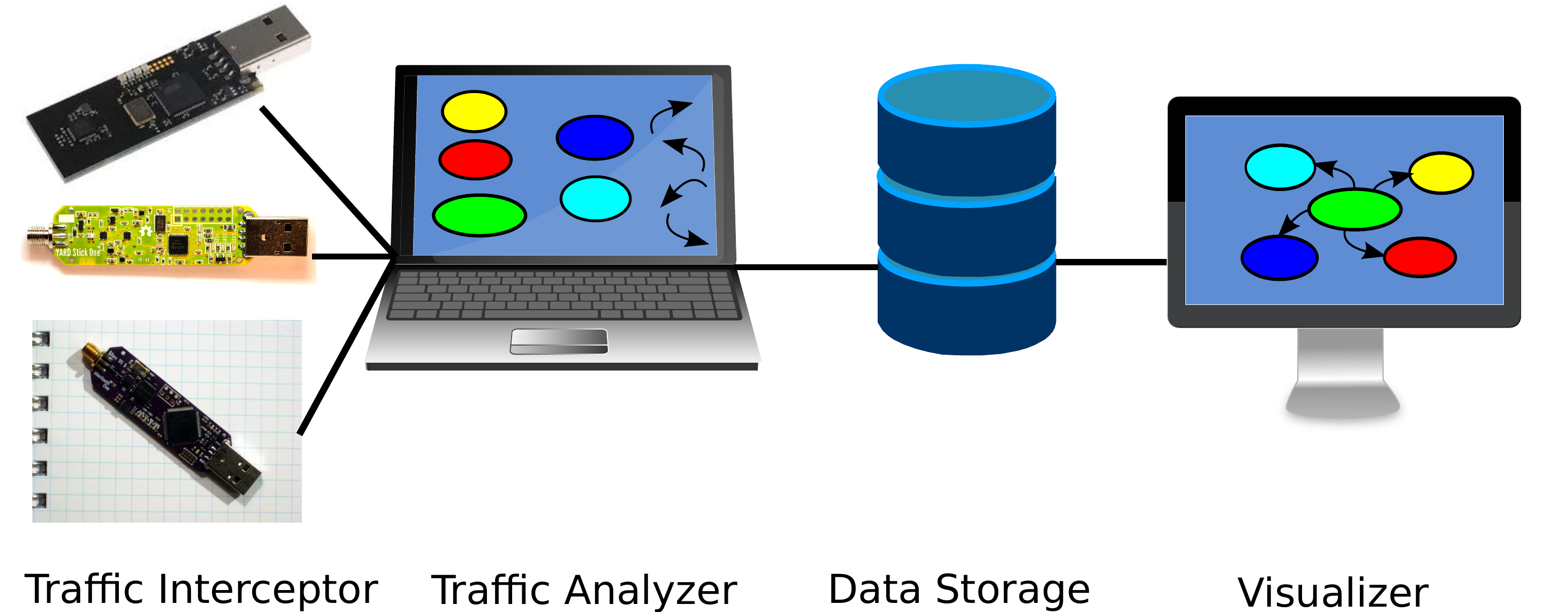}
	\caption{Overview of the IoTScanner architecture, which consists of four modules: traffic interceptor, traffic analyzer, data storage, and traffic visualization.} 

	\label{fig:architecture}
\end{figure}

Figure~\ref{fig:architecture} shows the overall architecture of our
proposed \emph{Passive Analysis Framework}, termed as IoTScanner.  The IoTScanner has four main functional modules: traffic interceptor (captures wireless signals), traffic analyzer (analyzes MAC frames), data storage (stores processed frames and results), traffic visualizer (displays the status of network).

Each of these modules is described in detail in the following. We
provide a detailed comparison of the IoTScanner with other related
projects in Section~\ref{sec:related}.

\subsection{Traffic Interceptor}
The traffic interceptor module provides flexible low level access to
the wireless medium.  In the context of IoT, widely used protocols are
802.11/WiFi, Bluetooth, Bluetooth LE, Zigbee, and Z-Wave; each being
prevalent in particular application area(s).  For example, the devices
accessing Internet primarily use WiFi, wearable/on-body devices (e.g.,
blood pressure monitor, smart watch) use Bluetooth or Bluetooth LE to
connect to their parent devices, smart home appliances (e.g., smart
meter) use Zigbee, and electronic appliances (e.g., AC, fridge, fan)
use Z-Wave protocols~\cite{atzori2010internet}. However, there is no
clear line of separation on what device can use which protocol, it
primarily depends on the usage of the devices or the energy
consumption behavior of them.  Therefore, it is important to have
interception capabilities leveraging multiple radios, either one radio
for each protocol, or software defined radios to cope up with the
variety of protocols available in the IoT spectrum. 

If multiple channels of the same standard should be reliably
monitored, it might even be necessary to use multiple radios of the
same kind, each on its own channel.

\subsection{Traffic Analyzer}
The traffic analyzer scrutinizes each Link layer frame, captured by the traffic
interceptor, by parsing only the frame overhead (i.e., header and
trailer) harnessing on passive analysis philosophy of the IoTScanner.
It extracts relevant information, such as the source and destination
addresses, frame type and sub-type, SSIDs present, from those the
frames to be used for targeted analytics.  We note that the frames
need to be treated on a protocol by protocol basis, because parsing
Bluetooth LE frames is significantly different from parsing WiFi
frames or Zigbee frames.  In addition, the analyzer records some additional
useful information such as the channel number on which the interceptor
is capturing traffic, size of the frame captured (in bytes) and the
timestamp at which the frame is captured.  It sends the extracted
frame information to the data storage module on per frame basis, where
actual analytics is performed.

\subsection{Data Storage}
The data storage module, acting as a simple server, provides a stable
storage unit primarily for storing processed information about
individual frames.  The storage can be done by various means, however we
prefer to employ a simple and light weight database system. In addition to storing the information, the system  provides a simple interface
for querying the database to carry out analytics.  For example, the
data storage interface can be realized using a set of RESTful APIs
that can easily accessed over Internet, and be used by third-party
applications.  The APIs are provided for the following functions:
storage and retrieval of frame information, generalized categorical queries on the DB, storage and retrieval of analysis results.

\subsection{Traffic Visualizer}
The traffic visualizer provides a visual representation of the
observed IoT environment at a particular time window.  A network,
i.e., an IoT environment, can be viewed from different perspectives,
starting from device-to-device connection to device or link specific
information to semantics of the underline, possibly collaborative,
activity performed by IoT establishment.
In this paper, we consider a mix of all these perspective at a high
level for an initial comprehension of the IoT environment where we
display a graphical view of the underlying network along with some
supporting description of the devices and the associated links, if
any.  We do not infer any application specific information, e.g., if
the IoT is used for maintaining room temperature by controlling AC,
rather our interest in this paper is to create an inventory of the
devices present and the amount of traffic generated by each of those
devices.
Because the devices need not be sending or receiving frames all the
time, the traffic visualizer needs to automatically and periodically
update to reflect any changes (in terms of both the devices and links)
in the IoT environment under surveillance.  Note that the visualizer
displays only those nodes, and accounts for only those links that are
present in the database currently.

\section{System Implementation}
\label{sec:implementation}
In this section, we present our implementation of the four
proposed IoTScanner modules (traffic interceptor, traffic analyzer,
data storage, and traffic visualizer). An overview of our
implementation is shown in Figure~\ref{fig:architecture}. We use
different wireless interfaces for the traffic interceptor, Scapy (a
python library) for the traffic analyzer, SQLite for the data storage,
and Javascript for the data visualization. The data storage uses
RESTful APIs to communicate with the analyzer and the visualizer
modules. The interceptor and analyzer components run on a
Raspberry Pi 3 device, the data storage can be run on the same device
or a server, and the visualization is displayed on an Android Tablet.

\begin{figure*}[tb]
	\centering
	\vspace{-0.0in}	
	\includegraphics[width=0.8\linewidth]{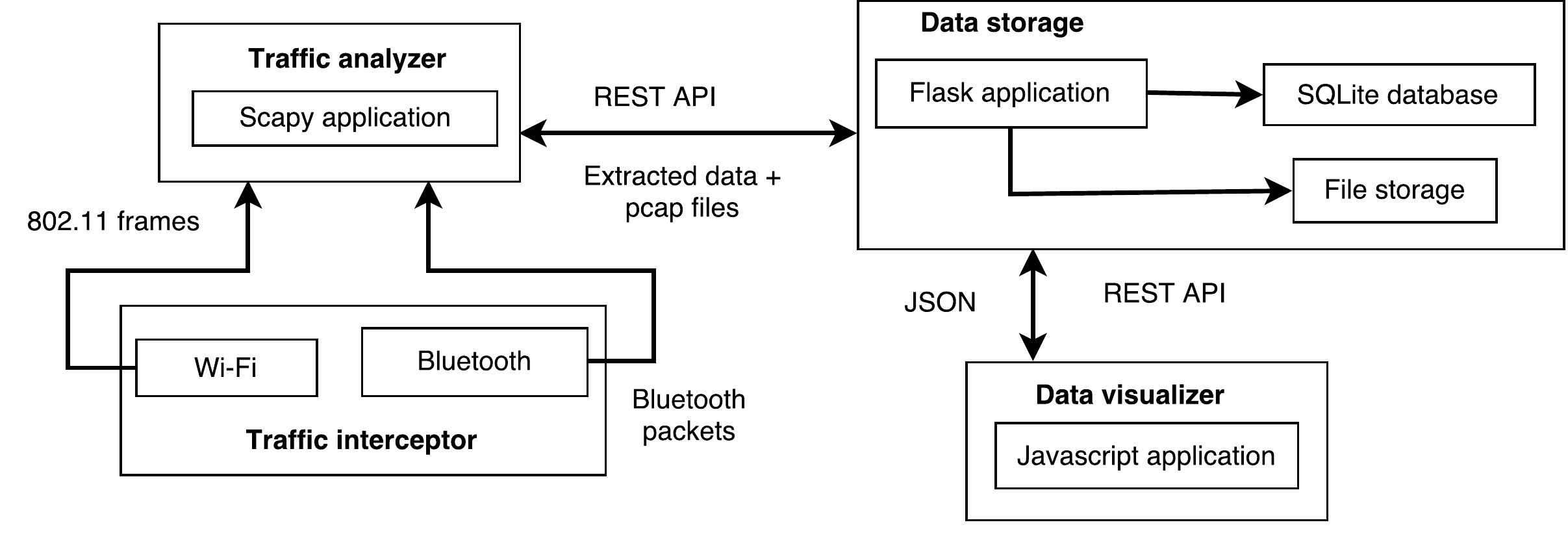}
	\vspace{-0.00in}
	\caption{Overview of IoTScanner implementation.}
	\label{fig:block_implementation}
\end{figure*}

\subsection{Traffic Interceptor}

The traffic interceptor module can be implemented
by a number of radio interfaces or by using software defined
radios. We decide to use the following radio interfaces that are commercially available, and connectable via USB: the TP-Link TL-WN722N 802.11n wireless
adapter (for WiFi), the Ubertooth One (for Bluetooth LE), and
Atmel-RZUSBstick (for Zigbee).

The TP-Link TL-WN722N adapter~\cite{tp-link} can be easily configured
to operate in monitor mode and capture WiFi frames with configurable
channel hopping (over 13 channels). The adapter can also support
certain active attacks such as deauthentication attacks, a useful
feature in case we extend the functionality of the IoTScanner to
include traffic decryption. We perform sequential channel hopping to
obtain an overview of the traffic on all channels.  The interface
dwells on a particular channel for a certain period of time before
hopping to the next channel. The dwell time can be configured by the
user, otherwise takes a default constant value.

We note that since we dwell on a single channel at a time, frames on
channels the interceptor is not listening on will not be captured. Hence, we capture a subset of the overall traffic.

For Bluetooth Low Energy (BLE) traffic capture, we use the Ubertooth
One~\cite{ubertooth}, an open source Bluetooth monitoring platform.
Compared to other Bluetooth monitoring solutions, this platform is
inexpensive and easily adaptable to our settings. We focus on
Bluetooth Low Energy (BLE) traffic as its presence is increasing in
IoT devices, especially those used in healthcare. The Ubertooth One is
able to detect the channel hopping map from sniffed BLE connection
request packets and follow connections by hopping in the same
pattern. The Ubertooth also has a promiscuous mode---to follow
connections that were already established at the time of sniffing. As
in the case of WiFi, the interceptor may miss out some devices as
result of channel hopping, which may be compensated by increasing the
observation period.

For Zigbee traffic capture, we use the RZUSBstick from
Atmel~\cite{rzstick} which supports low power wireless applications
using Zigbee, 6LoWPAN, and IEEE 802.15.4 networks.  This adapter can
also perform channel hopping (over 16 channels, from channel 11 to
channel 26 in the 2.4 GHz band). Other features of this traffic
capturing procedure is similar to WiFi or Bluetooth networks.

\subsection{Traffic Analyzer}
\label{subsec:analyzer}

\begin{figure}[tb]
	\centering
	\includegraphics[width=\linewidth]{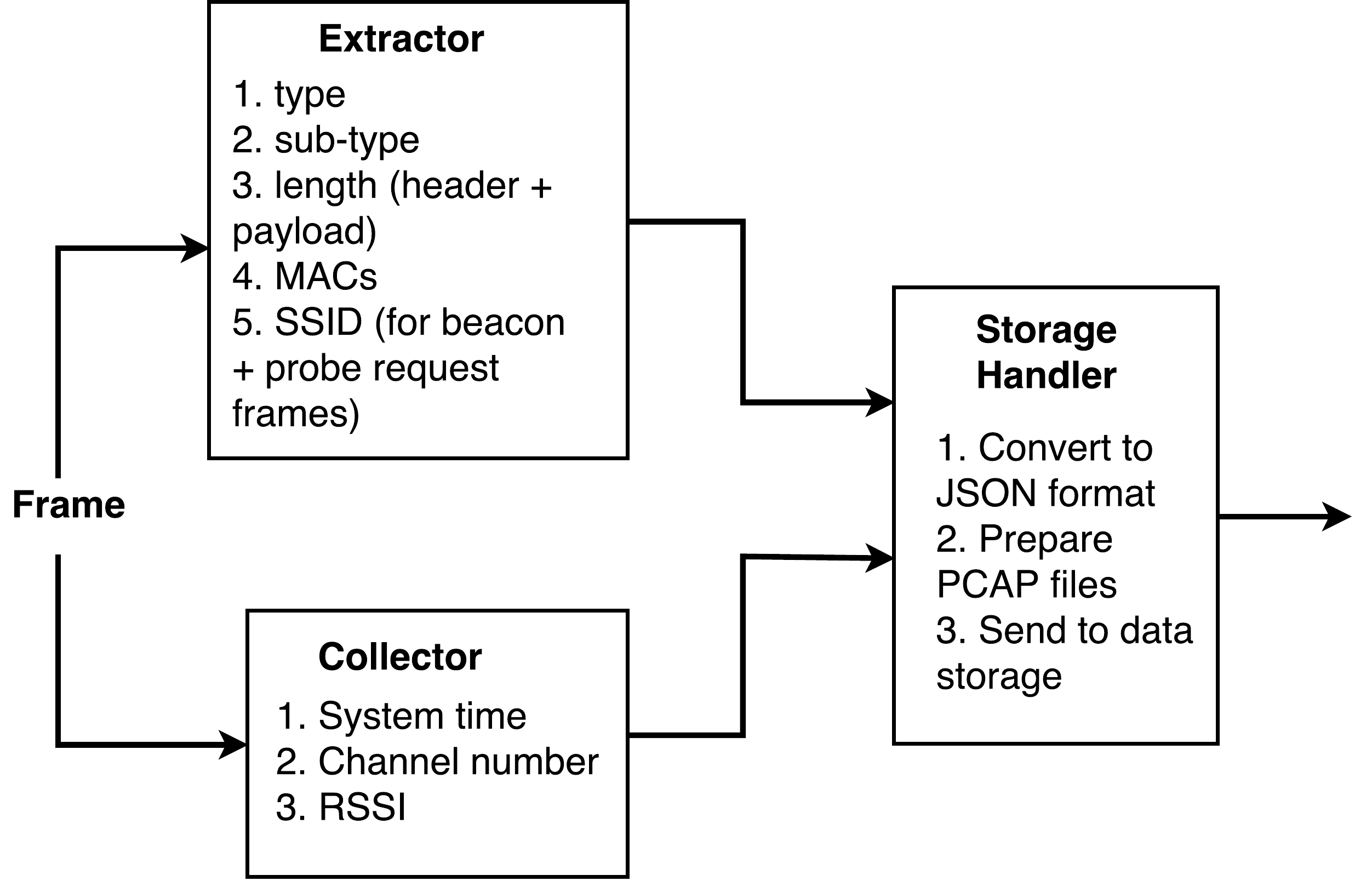}
	\caption{Traffic Analyzer module of our IoTScanner, for WiFi frames.}
	\label{fig:trafficAnalyzer}
\end{figure}

The traffic analyzer module consists of three sub-modules: extractor,
collector, and storage handler, as shown in
Figure~\ref{fig:trafficAnalyzer}.  The extractor and the collector
modules are implemented using Scapy~\cite{scapy}, a python library for
packet capture and analysis.  Every frame captured by the traffic
interceptor is an input to both the extractor and collector
sub-modules. Since the aim of the IoTScanner is to provide a quick
overview of the IoT environment, only relevant pieces of information
are extracted from every captured frame.  This is performed
\emph{online} (without buffering) so as to find quick answers to only
those questions that we have assumed to be sufficient to provide an
overview of the environment.  The extractor module extracts the
following from the respective frames:
\begin{itemize}
	\item In WiFi frames (type, sub-type, length, MAC address and SSID)
	\item In Bluetooth LE frames (type, length, MAC address type (public or random), MAC address, node local name)
	\item In Zigbee frames (type, length, PAN ID, addresses             )
\end{itemize}

The collector module collects information such as the system time during
frame capture, the channel number on which the frame is captured, and
the RSSI (for potential device localization).  Both of these modules
supply the captured information to the storage handler module.

The storage handler sub-module sends the collected and
extracted information to the data storage module.  It sends the frame information
(in JSON format) to the data storage module via HTTP POST method. 
The module also has the option of storing the frames in a PCAP file and periodically sending the files to the data storage for further analysis of the overall traffic (which might be more computationally intensive). 
It is worth mentioning that high level frame analysis is not possible
at the traffic analyzer since this requires aggregated information
from multiple frames, and hence such analysis is performed at the data
storage server end.

\subsection{Data Storage}
The data storage module of the IoTScanner provides a database server,
to be accessible via a set of APIs, to store and retrieve the
extracted/collected frame information.  Two modules, the traffic
analyzer and the traffic visualizer, interact with the storage module
using these APIs over network, potentially the Internet.  We have
implemented this module as a web server which interacts with a
light-weight database.  The APIs in the server are developed using
Flask (a Python web framework), which helps to easily build the
RESTful APIs for our purpose.  We use SQLite to build a light-weight
relational database system for our data storage
module.

\subsection{Traffic Visualizer}
We implement the traffic visualizer using Javascript with D3~\cite{d3} library for network visualization. 
The visualizer is compatible with any hand-held devices, such as smart phones, tablets, etc. in addition to the desktop browsers. 
The visualizer displays the IoT environment in a number of ways (e.g., summary text, connectivity graph, bipartite relation, etc.) to make it suitable for the user to understand different aspects of the underlying network.  
By default, it displays a network graph accompanied by a brief summary text.

\begin{figure}[tb]
	\centering
	\includegraphics[width=\linewidth]{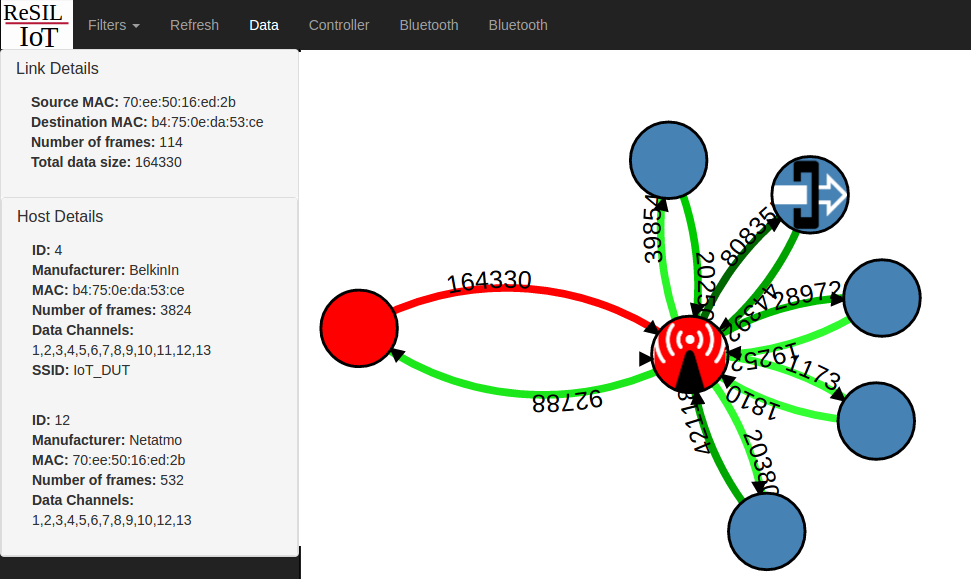}
	\caption{Example screenshot of our visualization app: IoT scenario represented using a graph structure. Selecting a node (red circle) in the graph displays the details about the underlying device.}
	\label{fig:DynamicGraphStruct}
\end{figure}

Figure~\ref{fig:DynamicGraphStruct} shows a sample network graph obtained during one of our experiments. The colored circles represent the nodes in the network and the arrows between a pair of nodes indicate that the pair exchanged at least one frame. We identify the access points from beacon and probe request frames and internet gateways using simple heuristics. The access points and gateways have icons to identify them in the visualizer.  

Our visual display is interactive in the sense that on selecting a node or an edge, more details about the selected component are displayed. 
For example, in the Figure ~\ref{fig:DynamicGraphStruct}, the selected nodes and edge are highlighted in red and their details are displayed in the sidebar. 
The graph is updated at a fixed interval of \emph{p} (configurable) seconds in order to reflect any changes in the number of devices or the communication links in the network. 

\begin{figure}[tb]
	\centering	
	\includegraphics[width=\linewidth]{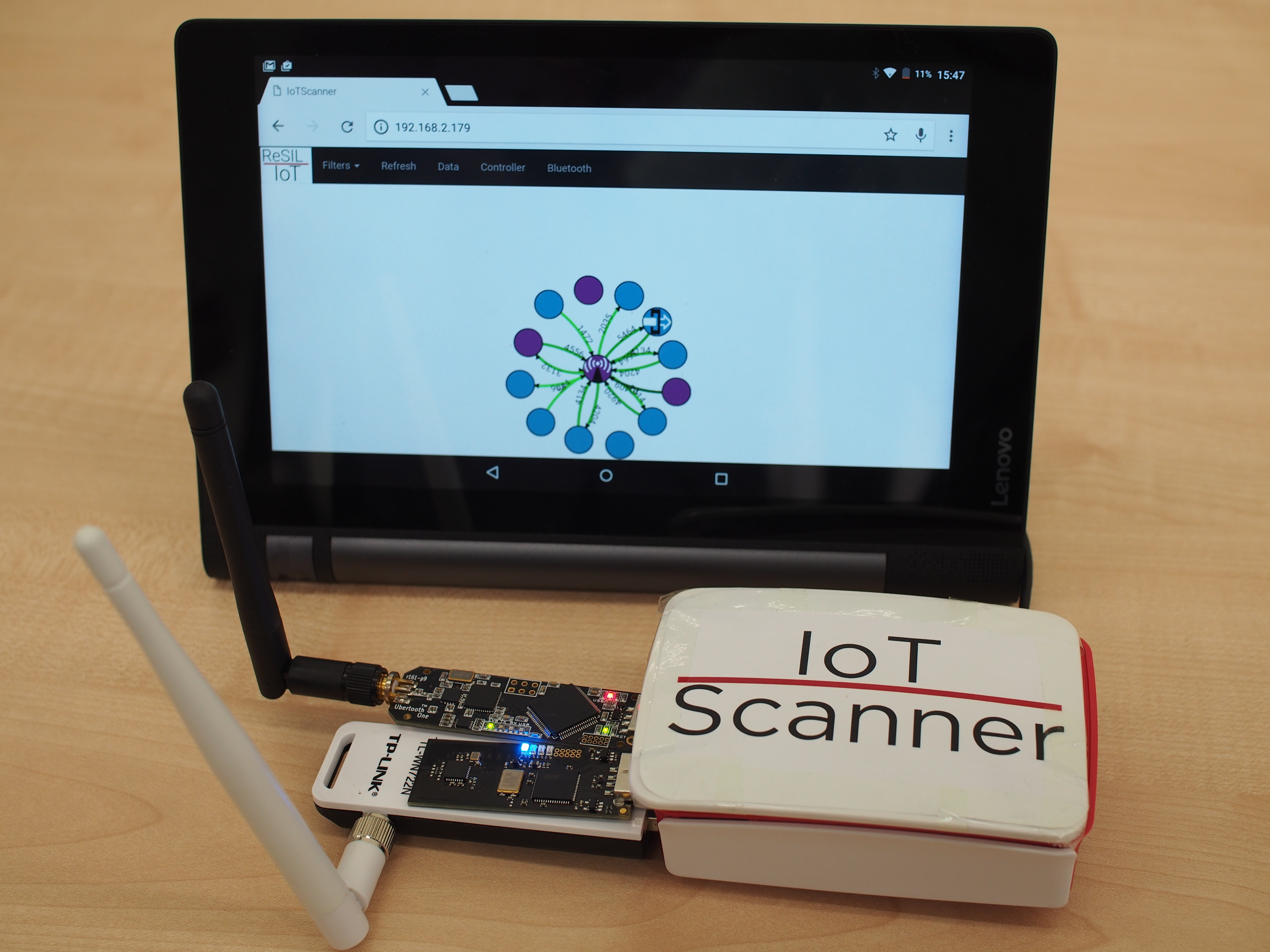}	
	\caption{IoTScanner with visualization on a hand-held device.}
	\label{fig:setup}
\end{figure}

Our overall implementation using the Raspberry Pi as interceptor (along with the tablet as visualizer) is shown in Figure~\ref{fig:setup}.

\section{WiFi Experiments}
\label{sec:wifi_exp}
In all our experiments, we place the scanner in an IoT testbed~\cite{siboni2016testbed} that contains a number of IoT devices. We conduct experiments with devices that use WiFi, Bluetooth Low Energy and Zigbee communication, with a larger focus on WiFi enabled devices since it is the predominant mode of communication for IoT devices. 
We perform our experiments in three phases after grouping the devices based on networking technology. 
In this section, we discuss the experiments using WiFi enabled IoT devices. Experiments with BLE and Zigbee are described in Sections ~\ref{sec:btevaluation} and ~\ref{sec:zigevaluation}.  

\subsection{IoTScanner configuration}

The IoTScanner, while intercepting WiFi traffic, uses two input parameters,
\begin{itemize}
\item Dwell Time ($T_d$): period of time (in seconds) that the traffic
  interceptor listens on a channel before moving to next channel
  ($T_d\in {5, 10, 20, 30 (default), 40, 50, 60}$).

\item Hops $T_h$: number of channel hops performed by the traffic
  interceptor ($T_h\in {1, 6, 13 (default), 26, 65, 130}$).
 
\end{itemize}

These parameters account for the amount of time the IoTScanner is
exposed to an IoT environment; for example, $T_h=13$ and $T_d = 5s$ implies
that the traffic interceptor scans for $13\times 5 = 65s$, and the
analysis will be performed only on the traffic captured during this period.

\subsection{Evaluation metrics}

Following are the common metrics for our experiments:
\begin{itemize}
\item \emph{nodes}: the total number of active devices in the observed
  environment (including access points). A device is considered to be
  active if it is observed to have sent/received at least one frame.
\item \emph{links}: the number of unique pairs of nodes that have
  exchanged at least one frame (excluding broadcast and multicast
  frames).

\item \emph{SSIDs}: the number of access points seen in the environment.
\item \emph{Frames}: the total number of frames (sent or received) per device; these are further classified by type into \emph{cFrames} (control), \emph{mFrames} (management) and \emph{dFrames} (data). 

\item \emph{Bytes}: the aggregated number of bytes (sent and received) per device; these are further classified by type into \emph{cBytes} (control), \emph{mBytes} (management) and \emph{dBytes} (data)
\end{itemize}

\subsection{Experiment Settings}
In our controlled experiments, we use six IoT devices: one Nest Cam
security camera, one Netatmo camera
(with face recognition feature), one TP Link security camera (of
relatively lower resolution compared to the other two cameras), one Amazon
Echo wireless speaker, one desktop with wireless adapter (to perform
general web surfing), and one WiFi access point. 
We conduct 10 experiments each, in a \emph{high-load} (being default setting) and a \emph{low-load} setting.

\Paragraph{High-load} This is the default setting for our
experiments, and in this setting, all three cameras (focusing on the same area) are
actively streaming video via Internet to a mobile device located outside the test
environment. The Amazon
Echo~\cite{echo} loudspeaker is streaming audio songs continuously
during the experiments. The desktop with wireless adapter is used to
browse web pages intermittently.

\Paragraph{Low-load} In this setting, all the devices are present but none of
them are actively used. For example, the IP cameras
are switched on, but the live video is not accessed. The Amazon
Echo is kept on but is not playing any music.

\Paragraph{Understanding the Network Structure}
First, we verify if IoTScanner can identify the nodes (with their MAC addresses) and the links among them from the captured traffic, hence determining the underlying network structure. 
We observe that our IoTScanner can correctly capture traffic from all the six devices in each of our experiments.
The scanner identifies the devices that sent out broadcast frames to advertise their presence in the network, and the wireless channels on which each of these devices sent/receive traffic. 
We experiment with various values (as noted earlier) of the two input parameters, dwell time ($T_d$) and hops ($T_h$). We observe that lower values of these parameters result in the scanner being unable to capture all the devices during the observation window. After multiple rounds of experiments, we conclude that $T_d=30$ and  $T_h = 13$ are optimal values to quickly capture a sufficient amount of traffic for the traffic classification analytics we want to perform. Finally, we use beacon and probe request frames to identify access points in the network, and simple heuristics (on amount of data and destination MAC addresses) to identify the Internet gateway. 

Results of these experiments are not presented here, as they are used more as a check of the correctness of our IoTScanner system. 
We present more interesting results that are obtained from the traffic analyzer of our system on device classification in an IoT environment.

\subsection{Per-node Traffic Classification} 
We perform simple analytics on the captured traffic to classify IoT devices. The mapping between the device labels we use and their actual name is shown in Table~\ref{table:wifidev}.

\begin{table}[t]
	\centering
	\begin{tabular}
          {l  l}
          \toprule 
          Label & Device Name\\
          \midrule
          Dev-1 & Desktop \\
          Dev-2 &    Netatmo camera\\
          Dev-3 &    TP-Link camera\\
          Dev-4 & Access Point \\	
          Dev-5 & Gateway \\
          Dev-6 & Amazon Echo \\
          Dev-7 &    Nest Cam camera\\
          \bottomrule		
	\end{tabular}
	\caption{Device labels and the corresponding devices used in WiFi experiments.}
        \label{table:wifidev}
\end{table}

\begin{figure}[tb]
	\centering
	\includegraphics[width=\linewidth]{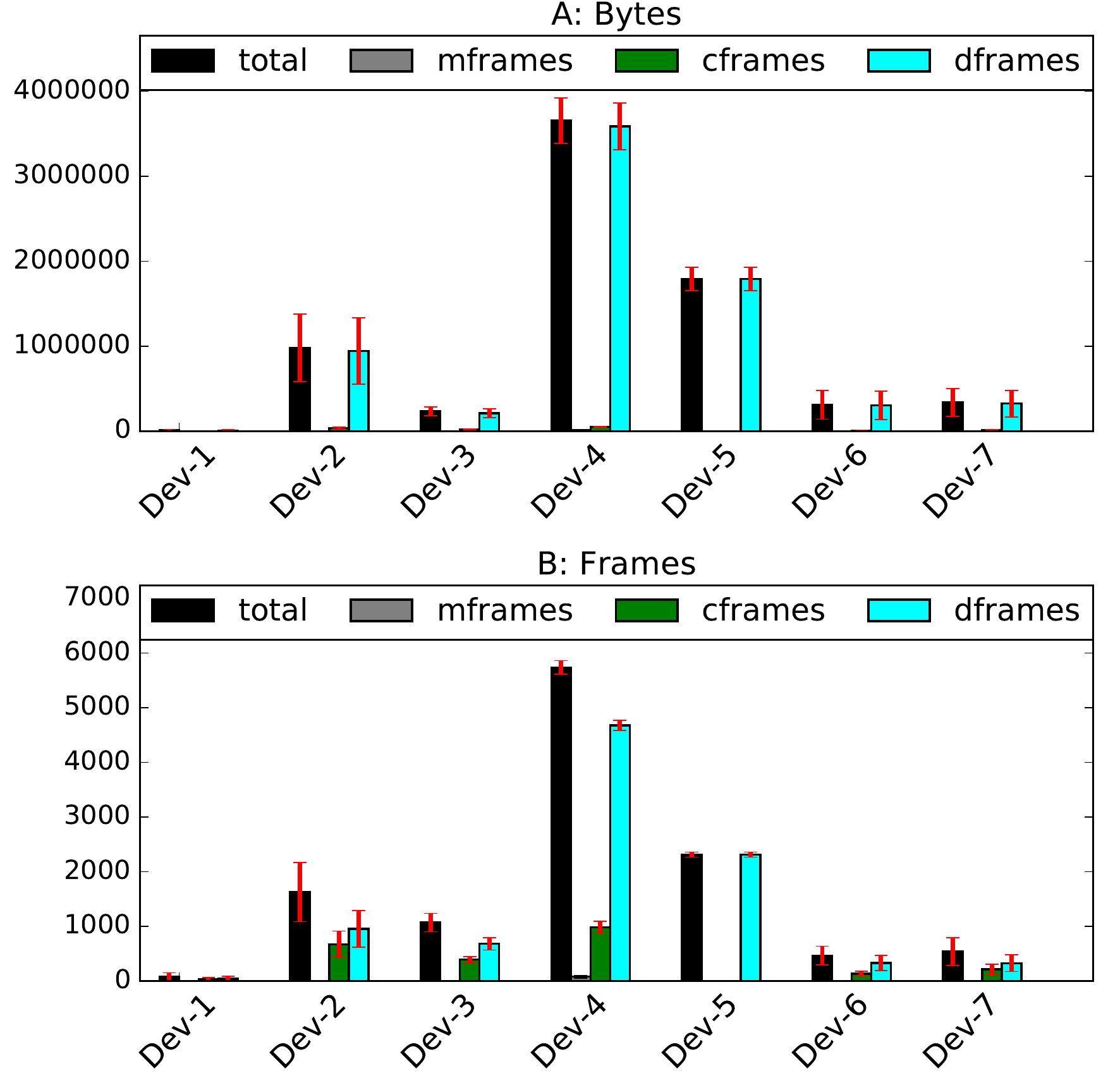}
	
	\caption{Variation of amount of traffic per device (subplot (A) in Bytes, and (B) in Frames) in high-load setting. }
	\label{fig:device_traffic}
\end{figure}

\Paragraph{Frames, mFrames, cFrames, and dFrames} 
First, we find the total amount of traffic associated with each device in the network, along with the type (management, control and data) in the high-load setting. We determine the traffic in terms of bytes (Figure ~\ref{fig:device_traffic}A) and frames (Figure ~\ref{fig:device_traffic}B). 
A frame is associated with a device if its MAC address is found in the frame either as the source or destination address. 

Interestingly, it can be seen that the traffic amount and its type can classify the devices at a high level. 
For example, the highest amount of traffic, in terms of both bytes and frames, is associated with Dev-4 which is the access point (see Table~\ref{table:wifidev}) that connects to all other devices present. 
Dev-5, which has no control and management frames, is the gateway device connected through Ethernet to the access point. 
The lowest amount of traffic is seen in Dev-1, the desktop, and it is used for occasional browsing during the experiments. 
The rest of the devices (Dev-2, Dev-3, Dev-6 and Dev-7) are associated with high traffic as they are either IP cameras or the Amazon Echo performing continuous streaming. 
In each of the devices, it can be seen that the data traffic (in bytes) dominates the control or management traffic, and is almost equal to the total amount of traffic of the corresponding devices. 
However, the difference between the number of control and data frames is not as high, especially in the case of the cameras. We observe that the acknowledgement frames contribute to the large number of control frames in this case. 

\begin{figure}[tb]
	\centering
	\includegraphics[width=\linewidth]{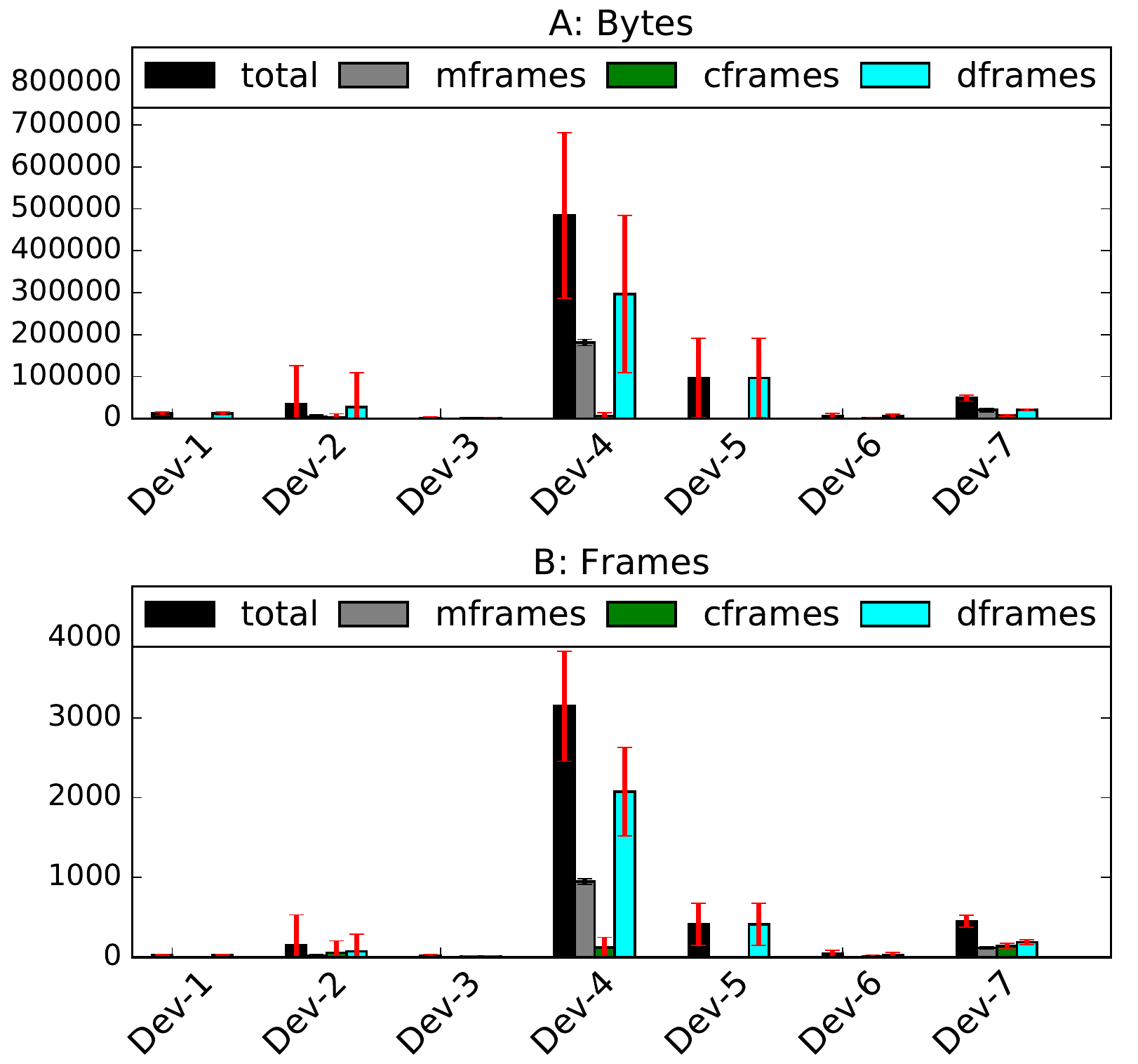}
	
	\caption{Variation of amount of traffic per device (subplot (A) in Bytes, and (B) in Frames) in low-load setting. }
	\label{fig:device_LowTraffic}
\end{figure}
We explore the traffic volume in low-load settings (results shown in Figure~\ref{fig:device_LowTraffic}).  
It can be seen that almost all the devices have control traffic comparable to data traffic (in bytes), as opposed to high-load settings where data traffic dominates the control traffic. 
In fact, standard deviation of traffic volume is quite significant in this setting in all the devices, perhaps because the devices send their status information more at times. 
Thus, an analysis of the traffic amount and its composition can potentially be used to learn if an IoT setting generates a high volume of data traffic. 

\begin{figure}[tb]
	\centering
	\includegraphics[width=\linewidth]{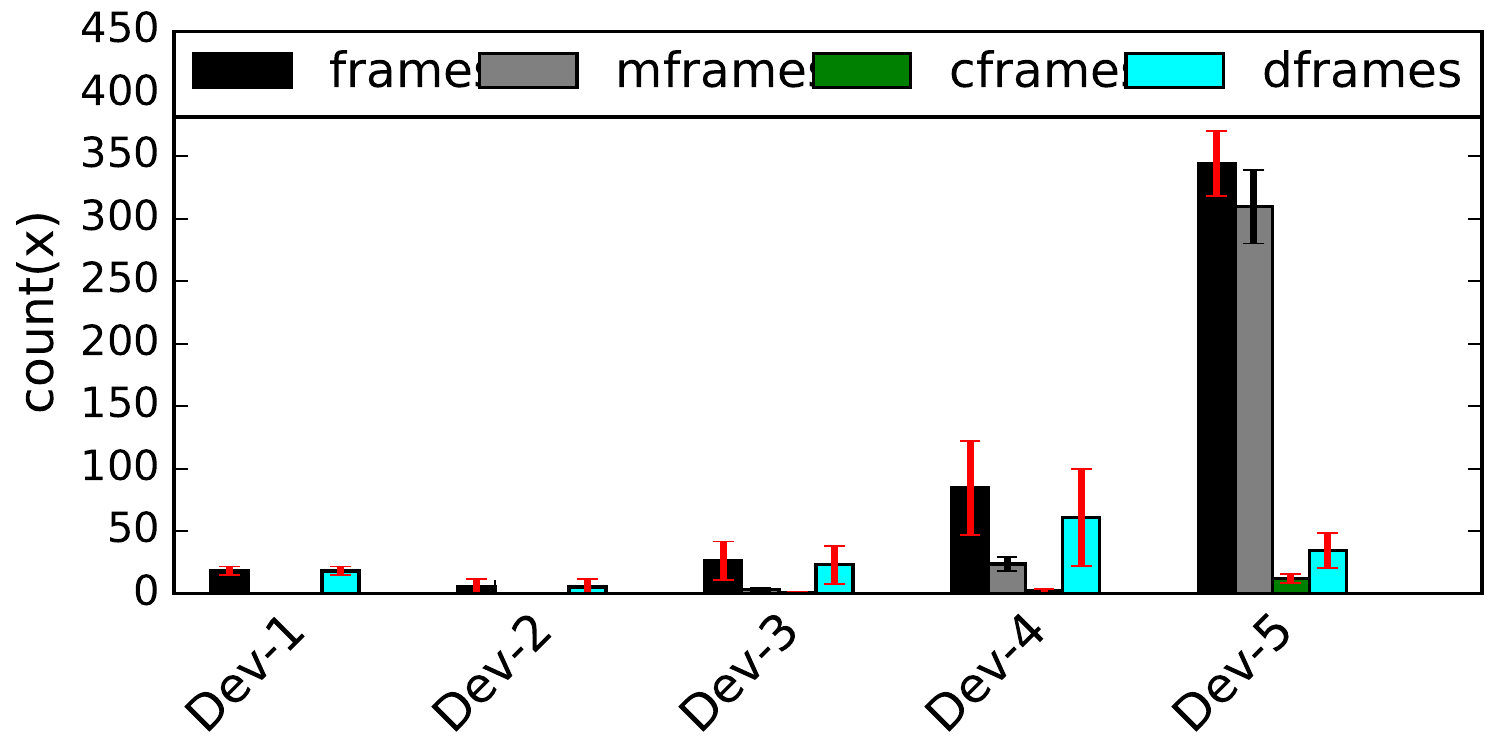}
	\caption{Variation of the number of frames and its types for each participating device in high-load setting.}
	\label{fig:validation_x_device}
\end{figure}

\Paragraph{Sent and Received Volume} 
Since the overall traffic mainly consists of data frames, we investigate the amount of data traffic (in terms of bytes and number of frames) sent and received by each device (Figure~\ref{fig:validation_x_device} shows results in high-load settings). The highest amount of traffic (either sent or received) is observed in \emph{Dev-4}, which is the access point. \emph{Dev-5} (the gateway) has about three times higher received traffic (in Bytes) as compared to sent. 
Note that the traffic towards the Internet is the received traffic for the gateway, and traffic coming into the local network accounts for sent for it. Thus, the result is consistent with the ground truth, as there are three cameras sending video traffic. We also notice that the cameras (\emph{Dev-2}, \emph{Dev-3} and \emph{Dev-7}) have a high amount of sent traffic, as expected. However, the amount of sent traffic varies significantly among the cameras. The received traffic is higher than the sent traffic for Dev-6, which is the Amazon Echo continuously streaming and playing audio songs from its server.
Our experiments indicate that active IoT devices such as IP cameras or music streaming devices can be identified by analysis of sent and received traffic volumes in high-load settings. 

\begin{figure}[tb]
	\centering
	\includegraphics[width=\linewidth]{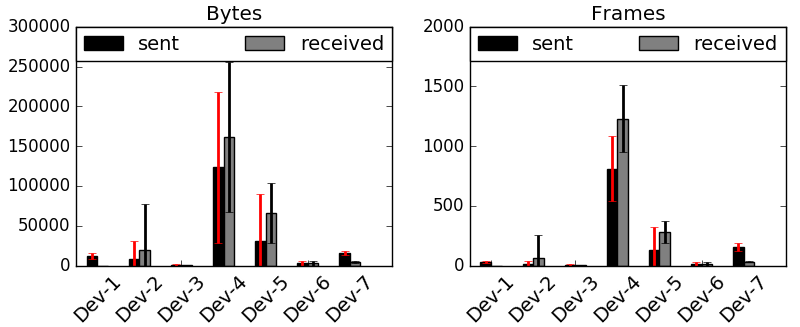}
	\caption{Variation of the number of frames and its types for each participating device in low-load setting.}
	\label{fig:validation_x_device_low}
\end{figure}

We also investigate traffic flow in the low-load settings (results shown in Figure~\ref{fig:validation_x_device_low}). 
Surprisingly, it can be seen that cameras do not necessarily produce a higher amount of sent traffic compared to receive traffic (e.g., Dev-2).
The Amazon Echo sends and receives almost equal amount of data in this setting.  
As expected, the gateway still receives more data than it sends, probably because these IoT devices continue to update their status to their associated cloud servers.

\begin{figure}[tb]
	\centering
	\includegraphics[width=\linewidth]{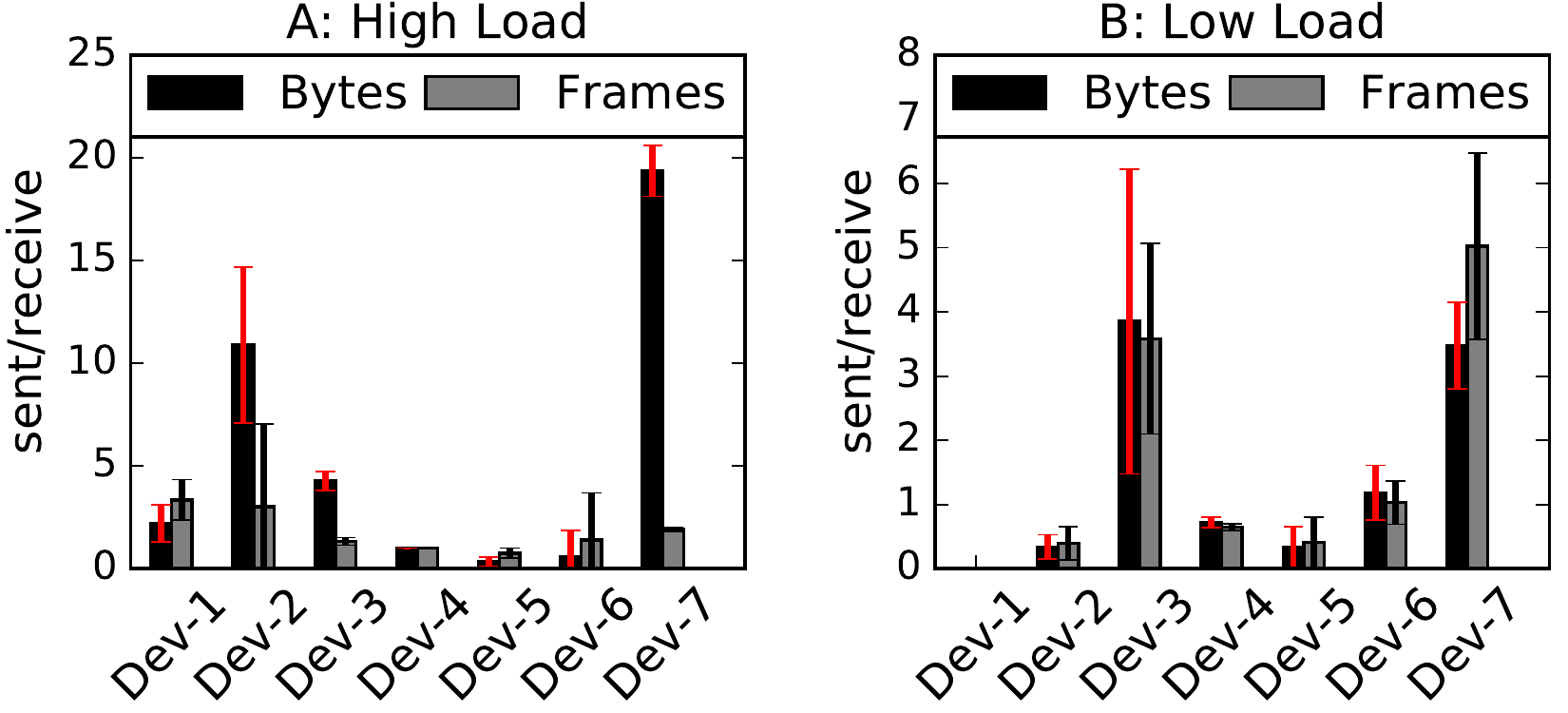}
	\caption{Variation of the ratio of sent and received traffic, per device basis in both high- and low-load settings.}
	\label{fig:privacy_x_device_y_sent_rec_high_low}
\end{figure}

\Paragraph{Sent-to-Received Ratio} We explore the possibility of identifying the devices by computing the ratio of sent to received traffic (in terms of both bytes and frames). We consider only data traffic for this analysis, and ignore management and control frames. Figures ~\ref{fig:privacy_x_device_y_sent_rec_high_low}A and~\ref{fig:privacy_x_device_y_sent_rec_high_low}B show the sent-to-received ratios in high-load and low-load settings respectively. In the high-load setting, the IP cameras (\emph{Dev-2}, \emph{Dev-3} and \emph{Dev-7}) have a ratio greater than 4 for traffic in bytes and greater than 1.5 for traffic in number of frames. This indicates that an IP camera that actively streams video traffic may potentially be identified when the ratio is greater than 1.0. Also, the ratio in bytes is greater than the ratio in frames for the cameras, implying that, per frame, a larger amount of data is originated from the cameras. 
The desktop with adapter (\emph{Dev-1}) has a lower ratio (>1.0) than the cameras but higher than the access point ($\approx 1.0$) and gateway (<1.0). The ratio of frames is much higher than the ratio in bytes, which indicates that the desktop sends more number of frames of smaller size. The access point can be clearly identified as it has a ratio close to 1.0 for both bytes and frames. The gateway (\emph{Dev-5}) has a ratio less than 1.0, which indicates higher received traffic. 
Finally, the Amazon Echo (\emph{Dev-6}) shows a ratio less than 1.0; as it continuously downloads audio traffic from the Internet. 

In the low load setting, the sent-to-receive ratio does not look promising as a metric to classify the devices. The ratio does not behave in the same manner for all the IP cameras - \emph{Dev-2} has a ratio less than 1.0, while \emph{Dev-3} and \emph{Dev-7} have a higher amount of sent traffic. The Amazon Echo has a ratio higher than 1.0 in this setting. Our experiments show that an analysis of the ratio alone in low-load settings may not be good enough to identify IoT devices.

\subsection{Classifying Streaming Camera}
Finally, we present a use case of our IoTScanner system that  addresses the issue of privacy in an IoT environment. In particular, we show how to differentiate nearby IoT cameras that are streaming data from other devices. 
To achieve that, we leverage our experience on traffic patterns as discussed in previous sections. It may be tempting to consider absolute traffic volume per device for the classification as any IP camera produces a lot of traffic compared to other devices (Figure~\ref{fig:device_traffic}). 
Unfortunately, there are a number of potential issues with that approach. The traffic volume can vary significantly depending on the observation window size. 
In addition, as different wireless channels can be used to carry traffic and our scanner performs channel hopping, it may not be able to capture all traffic for each and every device. 
The camera can also produce varying amount of traffic depending on its vendor (see Figure~\ref{fig:validation_x_device}).
Finally, the identification may become dependent on system parameters such as dwell time and hops.
Therefore, we consider ratio on different types of traffic volume as it looks promising (see Figure~\ref{fig:privacy_x_device_y_sent_rec_high_low}). 

We perform this analysis only on data traffic as it dominates control or management traffic in high-load settings for all the devices (Figure~\ref{fig:device_traffic}). 

We use two parameters here:
\begin{itemize}
	\item $R_{sr}$ - the ratio of sent to received data traffic, i.e., 
	$R_{sr} = Tr_s / Tr_r$, where $Tr_s$ and $Tr_r$ are the amount of data traffic (in Bytes) sent from and received respectively by a device. 
	\item $R_{bf}$ - the ratio of the traffic volume in bytes and in frames for a device, i.e., $R_{bf} = Bytes/Frames$ where Bytes and Frames are the amount of traffic in bytes and the frames respectively, as defined earlier. 
\end{itemize}

Our experimental setup consists of the three IP cameras as before and some additional WiFi enabled devices like smartphones, laptops, tablets, and a printer. We run eight experiments in total, having a varying number of active devices with at least one active camera in each experiment. We have a total number of 95 devices spread over all the experiments. 

The goal is to classify the cameras that are actively streaming. 
The status (active or non-active) of a camera is not changed during a single experiment though it may change across different experiments.

Results from the previous experiments give us an $R_{sr}$ range of $12 \pm 8$ and an $R_{bf}$ range of $1000 \pm 500$ bytes/frame. Thus, a device is identified as an actively streaming camera in an experiment only if it satisfies the conditions $4.0\leq R_{sr} \leq 20$ and $500 \leq R_{bf} \leq 1500$. 
In each of the experiments, we calculate the $R_{sr}$ and $R_{bf}$ for every device, and check if they can be identified as a streaming camera, otherwise it is identified as a non-camera device (denoted as ``others''). The results of these experiments are shown in a confusion matrix in Table~\ref{table:cmatrix}. We see that out of 95 device identifications, there were 3 false positives and 2 false negatives. This gives a false acceptance rate (FAR) of $\approx 3.61\%$ and a false rejection rate (FRR) of $\approx 16.67\%$. 

\begin{table}[t]
	\centering
	\caption{Confusion matrix to identify streaming camera, ``others'' = devices other than cameras. 	  \label{table:cmatrix}}
	\begin{tabular}{ r|cc }
 		$n=95$ & classified as camera & classified as others \\
		\cline{1-3}
		camera & 10 & 2 \\
		others & 3 & 80 \\
	\end{tabular}
\end{table}

False positive identification of one device occurs due to the presence of one Withings IP camera in our test settings. 
The camera is kept switched on and configured, but no remote access of live video is performed during the experiments. Hence, it is considered as a non-camera device. 
However, the camera starts streaming to its associated cloud server as soon as it detects some movement in the area of focus, causing it to be detected as an IP camera. 
Similarly, a couple of false negative cases are reported due to the Netatmo camera in our settings. This is because our scanner fails to sniff sufficient traffic for the device, as it performs channel hopping, making a false prediction case.

As the aim of our study is to defend against an ``honest but curious'' attacker that makes use of existing networking infrastructure to obtain insights about not only the IoT infrastructure itself but also human users associated with it, e.g., by accessing a surveillance camera, we use only MAC layer un-decrypted traffic for this purpose. 
Preliminary results of our analysis on identifying such streaming cameras in an IoT environment are promising and can lead to future studies for identifying other devices from MAC layer traffic obtained by passive sniffing. 
Thus, our IoTScanner can play an important role for initiating the process of addressing the problem of identifying potential privacy breaches in any personalized IoT environment. 

\section{Bluetooth LE Experiments}
\label{sec:btevaluation}
For the Bluetooth Low Energy (BLE) experiments, we consider six devices - a OnePlus smartphone, an August smart lock, two Fitbits, and two Lenovo tablets. 
The smartphone operates the smart lock via an Android app. 
Each tablet is associated with a Fitbit, and operates it via an Android app. 
In each experiment, one of the Fitbit is paired, allowed to sync its data and unpaired from the associated tablet; the smart lock is also paired, locked and unlocked (3-4 times) via commands from the app, and unpaired from the smartphone. 
Similar to the WiFi experiments, we conduct 10 experiments in each case, and compute an average of each of the measures. 
We do not attempt to decrypt the frames in these experiments as well.

We initiate experiments to
probe our network settings by detecting BLE nodes and links. Then, we
attempt to classify the devices via traffic categorization.  It is
seen that out IoTScanner is able to detect all three pairs of devices
and their links.  Surprisingly, we observe that the smart lock advertising frames do not follow BLE MAC address
randomization (i.e., a feature expected to be used by BLE devices for
privacy reason where the device MAC is replaced with some random MAC in
the BLE frames). In the Fitbit case, we do not detect advertising frames (refer to Section ~\ref{sec:discussion}), hence no information regarding MAC randomization is
revealed.  During the data exchange phase, the \emph{access address} is seen to
change every time the smart lock is paired. The Fitbit
pairs use a single access address across all pairing events.
Hence, the access address can be used to identify the Fitbit-tablet communication.

\begin{figure}[tb]
	\centering
	\includegraphics[width=\linewidth]{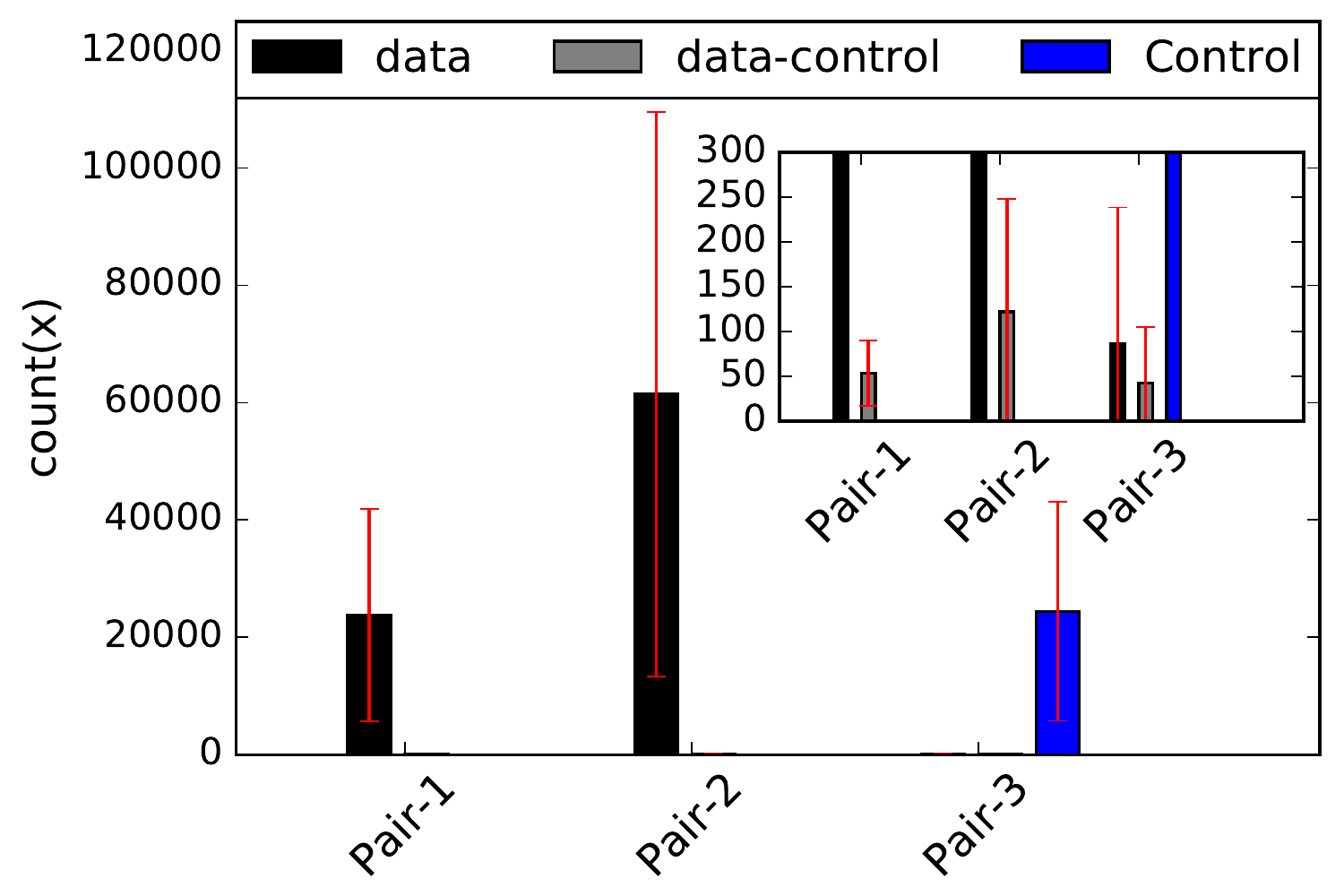}
	\caption{Variation of data and control frames (indicated by $x$ in \emph{count(x)}), Pair-1 and Pair-2 are Fitbit devices, and Pair-3 is smart lock device (inset plot shows data-control frames).}
	\label{fig:bluetoothExmt}
\end{figure}

In the smart lock case, we observe three types of control frames in addition to the advertising and data frames (BTLE\_DATA). Those are scan
request (SCAN\_REQ), scan response (SCAN\_RESP), and connection request
(CONN\_REQ) frames. No such control frames are seen in the Fitbit case. In both cases, we observe that the data frames (observed on
non-advertising channels) are further classified into data, control, and reserved sub-types. We denote the control sub-type frames as data-control (see Figure~\ref{fig:bluetoothExmt}) .

We observe traffic on all data channels in the case of the Fitbit, except for a few experiments where only one channel is seen to be carrying traffic. However, for the smart lock, only about 4 channels on average carry traffic. 

We present the results on the data and control traffic per BLE pair in Figure~\ref{fig:bluetoothExmt}. 

Both the Fitbit pairs generate significantly higher data traffic when compared to the smart lock (the inset figure shows a magnified version that displays data-control frames). 
This is expected as Fitbits perform a number of actions like movement detection, quality of sleep estimation and location information collection during every syncing with the phone, whereas the smart lock exchanges only lock/unlock commands. The variation in data traffic for the Fitbit can be explained by the differences in user activity for each experiment (relation between traffic and activity are described in ~\cite{das2016fitness}). 
Our experiments show that BLE pairs can also be classified using traffic volume analysis, though this is not done here due to insufficient number of devices at the time of experiments.

\section{Zigbee Experiments}
\label{sec:zigevaluation}

For the Zigbee experiments, we employ the Philips Hue lighting system that consists of one Hue bridge controlling three light bulbs. We intercept the traffic among the four devices and find the number of links, nodes and amount of traffic on each link. We conduct 10 experiments in total. In each experiment, we execute different commands on the three bulbs---one bulb is switched on and off once during the experiment, one bulb is switched on and off 6 times and one bulb is made to change its color 6 times. Each experiment lasts 120 seconds. The devices are labeled as shown in Table~\ref{table:zbdev}.

In all the experiments, our traffic interceptor detects all four devices (results are not presented here). 
We concentrate on the identification of devices using traffic analysis. 

\begin{table}[t]
	\centering
	\caption{Label-device mapping for Zigbee experiments.\label{table:zbdev}}
	\begin{tabular}
          {l  l}
          \toprule 
          Label & Device Name\\
          \midrule
          Zig-1 & Bulb - switched on and off once \\
          Zig-2 & Hue Bridge\\
          Zig-3 & Bulb - changed color \\
          Zig-4 & Bulb - switched on and off multiple times \\	
          \bottomrule		
	\end{tabular}
\end{table}

\begin{figure}[tb]
	\centering
	\includegraphics[width=\linewidth ]{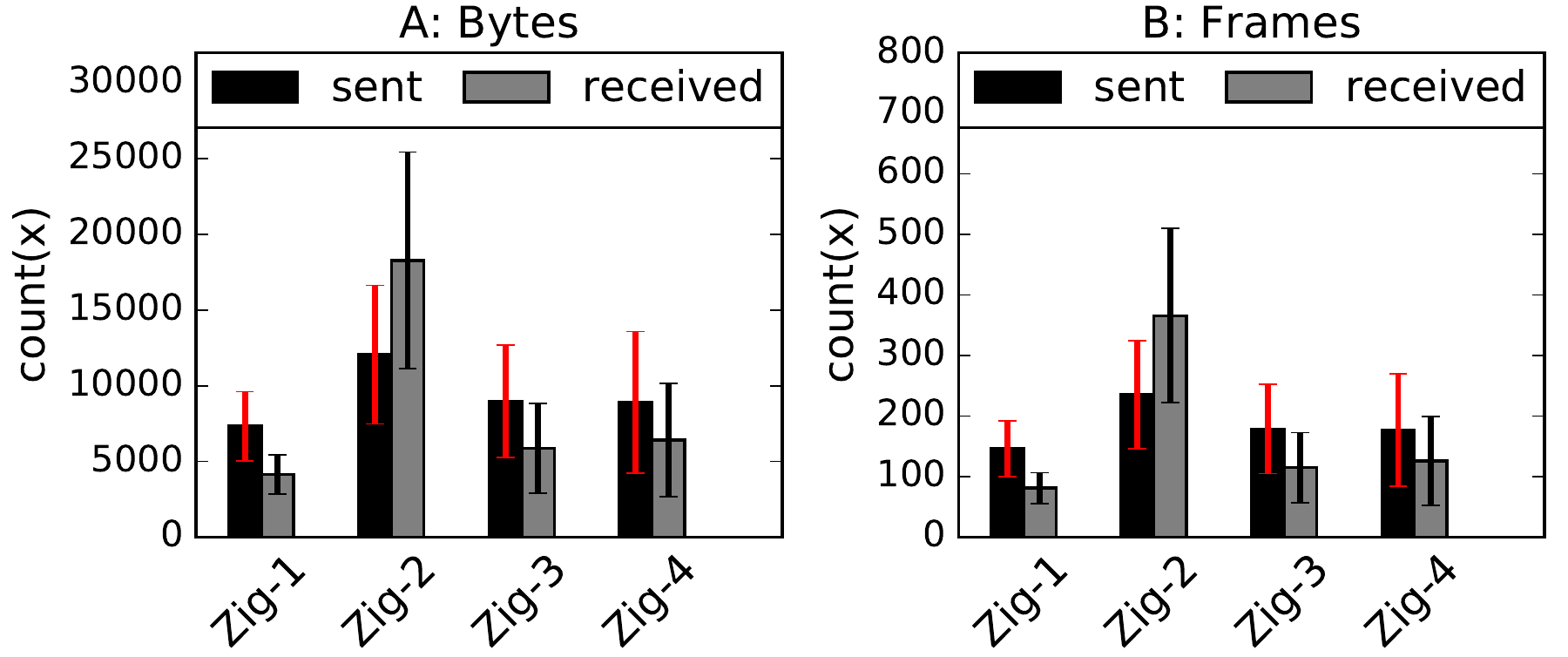}
	\caption{Variation of sent and received data (A) bytes and (B) frames of four Zigbee devices, one Philips Hue controlling three associated light bulbs.}
	\label{fig:zigbeeExmt}
\end{figure}

Figure~\ref{fig:zigbeeExmt} shows the amount of data sent and received by the four devices (each indicated by \emph{Zig-i}). We note that the Hue bridge (Zig-2) sends and receives a higher amount of data (in size and number of frames) when compared to the light bulbs. Among the light bulbs, the bulb with the lowest amount of activity (switching on and off once), i.e. Zig-1, has a smaller number of sent and received frames.   

The results also show that the controller receives about 40\% more data than it sends, whereas the bulbs have about 50\% less data received compared to sent. 
These initial tests show that an analysis of data sent and received can become a potential metric for device classification. It can be used to distinguish a light controller from its associated lights and might even help detect the amount of activity on the lights. 

\section{Discussion}
\label{sec:discussion}

In this section, we discuss some of such challenges here that we have faced during our experiments.

We observe that our wireless access point (LinkSys E1200) exposes two MAC addresses through the WiFi interface on certain occasions (for example, during our low-load experiments). On  scrutiny, we discover that these are the addresses of the WiFi and Ethernet interfaces. Our traffic analyzer module does not currently have a mechanism to correlate multiple MACs to the same access point. In addition, when we use the OnePlus smartphone to interact with the IP cameras, we observe that the IoTScanner detects multiple MAC addresses communicating with the camera despite the absence of other devices in the environment. Further investigation reveals that the Marshmallow Android OS used by the smartphone creates random MAC addresses to interact with the access point. Thus, it is important to devise a mechanism to detect multiple MAC addresses as belonging to a single device.  

During our BLE experiments, we observe that the smart lock uses standard advertising BLE channels 37, 38 and 39 to send control frames. Out of these, the connection request frame reveals the channel hopping sequence that is to be used during the data exchange phase. 
Therefore, the default ``follow connection'' mode of Ubertooth, which uses the connection request frame to follow connections, can be used to detect this BLE pair. However, we do not observe communication initiation on the usual advertising channels for the Fitbit. Hence, we had to use the ``promiscuous'' mode of the sniffer, which estimates the hopping sequence from frames on the data channels. This indicates the need to introduce pair specific sniffing policy in the case of BLE.

In addition, we notice that the Android applications of the BLE devices do not exhibit the same behavior. The smart lock application does not connect to the lock when the Bluetooth pairing is performed manually instead of through the application, probably indicating some security measures at the application level. However, the Fitbit application works even if the pairing/unpairing is performed manually during any Fitbit operation (such as syncing).
We do not observe similar issues while working with Zigbee devices.

\section{Related Work}
\label{sec:related}

In this section, we present work that is related to the problem of
monitoring complex wireless networks, including passive and active
sniffing systems. 

\Paragraph{WiFi monitoring} In~\cite{kotz2005analysis}, Kotz and Essien used
syslog messages, SNMP polling and tcpdump packet captures to
characterize WLAN usage on a college campus over a period of 77
days. Henderson et al.~\cite{henderson2008changing} built upon the
work of~\cite{kotz2005analysis} by capturing traces, including VoIP traces, from a larger set of access
points and users. In these works, the packets were captured by associating with access points and the trace analysis was done offline.

Davis developed a passive monitoring framework in~\cite{davis2004wireless}
to measure resource usage on 802.11b networks, and used it to analyze
various setups involving video streaming. Further work on resource
usage during streaming was done in~\cite{narbutt2006gauging}
and~\cite{narbutt2006experimental}. Yeo et al. implemented a wireless
monitoring system in~\cite{yeo2004framework}, using multiple sniffers
that produced a merged, synchronized trace which could be used for
Link layer traffic characterization and network diagnosis. They also
discussed the possibility of using anomalies in Link layer traffic for
security monitoring. The challenges posed by analyzing traces from
multiple sniffers was further explored in~\cite{mahajan2006analyzing}
and~\cite{cheng2006jigsaw}. In~\cite{mahajan2006analyzing}, the
authors introduced a finite state machine to infer missing packets
from a distributed system of sniffers. In~\cite{cheng2006jigsaw}, the
authors focused on large scale monitoring by utilizing 150 monitors to
capture 802.11 frames.  LiveNet~\cite{chen2008livenet} used multiple
sniffers to monitor sensor network deployments by reconstructing
routing behavior and network load from captured
traces. In~\cite{chen2008livenet}, the authors proposed algorithms for
route inference and topology reconstruction among nodes in a network
and provided visualization of the network topology and data
transfer. Chhetri and Zheng introduced the WiserAnalyzer
in~\cite{chhetri2009wiseranalyzer}---a passive monitoring tool to
capture wireless traces and infer relationships among nodes in the
network. Our proposed IoTScanner aims to provide similar visualization but in
real-time and for a larger number of protocols. In comparison to these
papers, where analysis of the collected data was done offline, our
work focuses on real-time passive analysis.

A real-time passive monitoring framework was developed by Benmoshe et
al.~\cite{benmoshe2011joint} and deployed on a university
campus. Details such as number of clients, channel, error rate
etc. were stored in a database and a map of active devices was built.
They provided some visualization in the form of a map of active
devices. In contrast to~\cite{benmoshe2011joint}, our work does not
have prior knowledge of the network setup. SNAMP~\cite{yang2006snamp}
was a multi-sniffer and multi-visualization platform for wireless
sensor networks that could perform capture and visualization in
real-time. Our work aims to enhance some of the features mentioned in
these systems, with the introduction of APIs for easy access to data,
more visualization features and monitoring of other protocols
(Bluetooth and Zigbee).

Kismet~\cite{kismet} is one of the most widely used real-time, passive sniffing tools. It is targeted at monitoring 802.11 networks but offers plugins for Bluetooth and Zigbee traffic capturing. Though it provides some analysis in terms of enumerating the wireless networks, hosts and amount of data it sees, higher level analysis has to be done manually. In addition to this, Kismet does not have a detailed visualization tool. Some tools have been built on top of Kismet, mainly for the purpose of visualizing the node locations (using GPS plugins) but several of them are no longer maintained and use outdated libraries. Wireshark~\cite{wireshark} is also a well known tool for passive analysis. Like Kismet, it has plugins to handle Bluetooth and Zigbee captures. However, it does not have the provision for automated, detailed analysis or visualization of the observed network. The IotScanner shares many features with Kismet and Wireshark but hopes to enhance the user experience with more analysis and visualization tools. 
We compare a number of features of our system with related existing tools mentioned above. An overview of the features present in the IoTScanner and other tools is shown in Table~\ref{table:compare}.

\Paragraph{Fingerprinting} Franklin et al.~\cite{franklin2006driver} performed passive fingerprinting of 802.11 drivers by analyzing the durations between probe request frames sent by different drivers. This approach was fine-tuned in ~\cite{desmond2008identifying} to distinguish different operating systems using the same driver.  Pang et al.~\cite{pang2007802} identified four metrics that would help identify users from a network trace, out of which three could be used even with link layer encryption. These works use different metrics from ours for classification. Frame size is one of the factors in ~\cite{pang2007802} but they investigate this only for broadcast frames, while our work focuses on size and directionality of data frames.

\Paragraph{BLE monitoring} Spill and Bittau~\cite{spill2007bluesniff} developed an open-source single channel Bluetooth sniffer, BlueSniff, that could discover the MAC addresses of Bluetooth devices. The Ubertooth has a larger set of features at lower cost when compared to BlueSniff.  
Ryan~\cite{ryan2013lowenergy} built a tool to sniff Bluetooth Low Energy (BTLE) communication, on the Ubertooth platform. This tool is also able to follow connections that were already established at the time of sniffing. He implemented a traffic injector to demonstrate passive attacks against 
BTLE. The BTLE sniffing is now part of the Ubertooth project and we utilize it for the IoTScanner platform. Albazrqaoe et al.~\cite{albazrqaoe2016blueear} presented the BlueEar system, a sniffer that is also based on Ubertooth but tries to identify and improve the factors that degrade sniffing performance. They also discuss the implications of privacy leakage in BLE devices and the importance of developing tools for BLE sniffing. While the focus of the BlueEar system is on sniffing Classic Bluetooth, our work deals with BLE communication. Privacy in BLE communication of fitness trackers was explored in ~\cite{das2016fitness}. They used the local name parameters in BLE advertising packets to detect fitness trackers. They also discovered that fitness trackers send a large number of advertising packets and do not randomize their MAC addresses, making them vulnerable to tracking. They analyzed data packets from Fitbits and concluded that the volume of data sent can be correlated to the level of user activity. 

\Paragraph{Zigbee monitoring} There are a number of works that discuss attacks on the the Zigbee Light Link standard. A few examples are ~\cite{muller2016security}, ~\cite{dhanjani2013hue} and ~\cite{morgner2016all}. However, the focus of these is more on active attacks. Dos and Lauradox ~\cite{dos2015preserving} explored a passive approach where they used the killerbee framework and Wireshark to sniff communication in a Zigbee mesh network. Their motivation is similar to the IoTScanner since they also aim to construct the topology of the Zigbee network and find privacy leaks. However, their work was conducted on a platform of motes they developed, while we focused on analyzing commercial IoT devices.

\begin{table}[t]
	\centering
	\caption{Proposed IoTScanner Features vs Related Works. $\Tdot=$ supported.\label{table:compare}}
	\begin{tabular}
          {l @{}c@{} @{}c@{} @{}c@{} @{}c@{} @{}c@{} @{}c@{} @{}c@{} @{}c@{} @{}c@{} @{}c@{} @{}c@{} @{}c@{}}
          \toprule 
          Related Work&\rota{Passive}&\rota{Real-time}&\rota{Autom. Analys.}&\rota{WiFi}&\rota{Bluetooth}&\rota{Zigbee}&\rota{API}&\rota{Visualization}\\
          \midrule
          Proposed work & \Tdot & \Tdot & \Tdot & \Tdot & \Tdot & \Tdot & \Tdot & \Tdot \\
          Wireshark~\cite{wireshark} & \Tdot  & \Tdot &  & \Tdot & \Tdot & \Tdot & &  \\
          Kismet~\cite{kismet} & \Tdot & \Tdot &   & \Tdot & \Tdot &  \Tdot & & \Tdot \\
          Benmoshe et al.~\cite{benmoshe2011joint} & \Tdot & \Tdot &   & \Tdot & &  & & \Tdot  \\	
          Chhetri et al.~\cite{chhetri2009wiseranalyzer} & \Tdot &  &  & \Tdot  &  & & &  \\
          Yang et al.~\cite{cheng2006jigsaw} &  \Tdot &  \Tdot & \Tdot & \Tdot &  &  &  &  \Tdot \\
          Chen et al.~\cite{chen2008livenet} & \Tdot &  & \Tdot  & \Tdot &  &  & & \Tdot \\
          \bottomrule		
	\end{tabular}
\end{table}

\section{Conclusion}
\label{sec:conclusions}

In this paper, we proposed the IoTScanner, a system that can passively
scan and analyze an IoT environment. We described the architecture and
implementation details of our system. The IoTScanner currently scans
WiFi, Bluetooth Low Energy and Zigbee traffic. It can analyze the
traffic to provide an overview of the devices that are active in the
environment and the communication among them.  We conducted
experiments to evaluate the performance of the IoTScanner in WiFi, BLE
and Zigbee environments. 

We introduced a simple ratio analysis of sent-to-receive traffic that
can classify WiFi enabled devices in an active IoT environment, which
can potentially contribute to the privacy related issues of a
personalized IoT environment.  In our experiments, we showed that we
were able to identify actively streaming cameras in our
environment. Out of 95 device classifications, our system had 3 false
positives (FAR=3.61\%) and 2 false negatives (FRR=16.67\%).

In future work, we aim to incorporate more types of IoT devices into
our experiments and introduce advanced analytics to classify the
devices based on link layer traffic. We are also interested in
enhancing the BLE and Zigbee capabilities of the IoTScanner.

\balance 
\bibliographystyle{abbrv}
\bibliography{bibs/bibliography}

\end{document}